\documentclass[a4paper,aps,prd,10pt,preprintnumbers,showpacs,twocolumn,superscriptaddress,nofootinbib,amsmath,amssymb,gensymb,amsfonts]{revtex4-1}
\usepackage{graphicx}
\usepackage{amsmath,amssymb,bm}
\usepackage{epstopdf}

 \def\imo{i}

\def\imo{i}

\def\be{\begin{equation}}
\def\ee{\end{equation}}
\def\bea{\begin{eqnarray}}
\def\eea{\end{eqnarray}}

\def\imo{i}
\def\re#1{Re(#1)}
\def\im#1{Im(#1)}
\def\K{{\cal K}}
\def\Order#1{{\cal O}\left(#1\right)}

\begin{document}

\title{Correspondence between quasinormal modes and grey-body factors of spherically symmetric traversable wormholes}
\author{S. V. Bolokhov}\email{bolokhov-sv@rudn.ru}
\affiliation{Peoples' Friendship University of Russia (RUDN University), 6 Miklukho-Maklaya Street, Moscow, 117198, Russia}
\author{Milena Skvortsova}\email{milenas577@mail.ru }
\affiliation{Peoples' Friendship University of Russia (RUDN University), 6 Miklukho-Maklaya Street, Moscow, 117198, Russia}

\begin{abstract}
A correspondence between two distinct spectral problems, quasinormal modes and grey-body factors, has recently been established for a wide class of black holes. Here, we demonstrate that a similar correspondence exists for a broad class of traversable wormholes and verify it using several well-known examples.
\end{abstract}

\pacs{04.30.Nk,04.50.+h}
\maketitle

\section{Introduction}

The observation of wormholes, theoretical passages through spacetime connecting distant regions or universes, remains a compelling frontier in modern physics and astrophysics. Rooted in solutions to Einstein's field equations, such as the Einstein-Rosen bridge, their detection would require indirect evidence through phenomena like gravitational lensing, unique signatures in quasinormal modes, or anomalous energy distributions inconsistent with black hole models. Advanced tools like high-resolution interferometers, precise timing of astrophysical signals, and multi-messenger astronomy could identify such signatures. However, observational challenges are compounded by the need for exotic matter to sustain traversable wormholes, stringent constraints from stability analyses, and the rarity of potential wormhole-hosting environments. Nonetheless, their discovery would revolutionize our understanding of spacetime topology and cosmic connectivity, offering unprecedented insights into quantum gravity and the fundamental nature of the universe.

Wormholes, as exotic objects, can emulate the ringdown phase of Einsteinian black holes \cite{Damour:2007ap,Cardoso:2016rao}
Their quasinormal spectrum offers diverse interpretations \cite{Aneesh:2018hlp,Bueno:2017hyj,Maselli:2017tfq,Nandi:2016uzg,Konoplya:2010kv}, while observations of optical phenomena in the electromagnetic spectrum do not exclude their existence \cite{Zhou:2016koy,Tsukamoto:2017hva}. 

Despite not being ruled out by current astrophysical data, wormholes pose significant theoretical challenges concerning their viability. Here, we focus on Lorentzian traversable wormholes described by the Morris-Thorne ansatz \cite{Morris:1988cz}, assuming symmetry about their throats and disregarding possible echo effects. This symmetry leads us to effective potentials with a barrier-like form, symmetric about the throat, where the potential's maximum aligns with the throat's position. Such behavior is typical for many Morris-Thorne metrics with a monotonic red-shift function in terms of the radial coordinate $r$ and is expected for other massless neutral test fields in the high multipole number ($\ell$) regime \cite{Morris:1988cz}.

A fundamental requirement for the viability of a wormhole model is its stability under small spacetime perturbations. The stability of thin-shell wormholes has primarily been examined for purely radial perturbations \cite{Poisson:1995sv,Lobo:2003xd,Eiroa:2003wp}, but this analysis remains inconclusive, while many other wormhole solutions have been shown to be unstable. \cite{Bronnikov:2012ch,Cuyubamba:2018jdl,Cremona:2018wkj}.
In this work, we set aside such existential questions and adopt an agnostic perspective, assuming the existence of a stable wormhole whose high-frequency radiation follows the standard eikonal regime.

The correspondence between quasinormal modes and grey-body factors has recently been established for spherically symmetric, asymptotically flat, or de Sitter black holes \cite{Konoplya:2024lir} and has been further extended to certain rotating black holes \cite{Konoplya:2024vuj}. Notably, grey-body factors may exhibit greater stability against deformations of the near-horizon geometry \cite{Oshita:2023cjz,Rosato:2024arw} compared to the overtones of quasinormal modes, which are known to be highly sensitive to such deformations \cite{Konoplya:2022pbc,Konoplya:2022hll}.  Consequently, the correspondence between quasinormal modes and grey-body factors have been used to study the letter in \cite{Dubinsky:2024vbn,Skvortsova:2024msa}.

The correspondence was found for black holes using the WKB approach to both characteristics. 
Thus, it is not clear how universal it is.  In the eikonal limit the correspondence looks as the exact one, because this is the limit where the WKB result is exact. Nevertheless, it is worth noting that in several theories incorporating higher curvature corrections or cosmological constant, the eikonal quasinormal spectrum of gravitational perturbations deviates significantly from the standard behavior observed for test fields \cite{Konoplya:2017wot,Konoplya:2017lhs,Konoplya:2022gjp,Bolokhov:2023dxq}.   However, for the aforementioned examples when the WKB method does not work or describes only part of the high frequency spectrum \cite{Konoplya:2017wot,Konoplya:2017lhs,Konoplya:2022gjp,Bolokhov:2023dxq} the correspondence does not work or is related only to the WKB part of the spectrum respectively.

Given that the boundary conditions for quasinormal modes and grey-body factors are the same for both black holes and wormholes, it is reasonable to conjecture that a similar correspondence could exist for asymptotically flat or de Sitter traversable wormholes. In this paper, we demonstrate that this is indeed the case.

While the literature on quasinormal modes of wormholes is quite extensive (see, for instance, \cite{Bronnikov:2021liv,Churilova:2019qph,Bronnikov:2019sbx,Batic:2024vsb,Churilova:2021tgn,Blazquez-Salcedo:2018ipc,Oliveira:2018oha,DuttaRoy:2019hij,Ou:2021efv,Azad:2022qqn,Zhang:2023kzs,Alfaro:2024tdr,Azad:2023idq,Mitra:2023yjf,Gogoi:2022ove,Roy:2021jjg,Jusufi:2020mmy} and references therein), the calculation of grey-body factors has been carried out only for a few specific cases \cite{Konoplya:2010kv}. 

In this work, we will examine several well-known examples of traversable wormholes, compute their grey-body factors, and analyze their correspondence with quasinormal frequencies, both in the eikonal limit and beyond.

Our paper is organized as follows: In Sec.~II, we briefly introduce the Morris-Thorne ansatz for wormholes. In Sec.~III, we derive the correspondence between quasinormal modes and grey-body factors for wormholes and discuss the WKB method used for these calculations. Sec.~IV is dedicated to exploring various examples of wormholes to test this correspondence. Finally, in the Conclusions, we summarize the results obtained.

\section{Quasinormal mode problem for traversable wormholes}

Static, spherically symmetric, Lorentzian traversable wormholes of arbitrary geometry can be described using the Morris-Thorne framework \cite{Morris:1988cz}. The spacetime metric is given by:

\begin{equation}\label{MT}
ds^2 = - e^{2 \Phi(r)} dt^2 + \frac{dr^2}{1 - \frac{b(r)}{r}} + r^2 (d\theta^2 + \sin^2\theta \, d\phi^2).
\end{equation}

Here, $\Phi(r)$ represents the lapse function, which governs the redshift and tidal forces within the wormhole spacetime. If $\Phi(r)$ is constant, the wormhole is referred to as "tideless." According to the terminology introduced in \cite{Morris:1988cz}, a tideless wormhole does not induce acceleration on point particles. However, tidal forces on extended objects may still persist under these conditions. The geometry of the wormhole is shaped by the function $b(r)$, known as the shape function.

The throat of the wormhole is located at the minimum value of the radial coordinate, $r_{\text{min}} = b_0$. The radial coordinate $r$ ranges from this minimum value at the throat to spatial infinity, $r \to \infty$. To better understand the geometry, we use the proper radial distance coordinate, $l$, defined by:

\begin{equation}
\frac{dl}{dr} = \pm \left(1 - \frac{b(r)}{r}\right)^{-1/2}.
\end{equation}

In terms of $l$, the spacetime extends symmetrically from $l = -\infty$ to $l = +\infty$, corresponding to $r \to \infty$ on either side of the wormhole.

To ensure the absence of singularities in the spacetime, the lapse function $\Phi(r)$ must remain finite everywhere. Furthermore, asymptotic flatness requires that $\Phi(r) \to 0$ as $r \to \infty$ (or equivalently $l \to \pm \infty$). The shape function $b(r)$ must satisfy the following conditions to ensure the wormhole's traversability and physical viability:

\begin{equation}
1 - \frac{b(r)}{r} > 0, \quad \frac{b(r)}{r} \to 0 \quad \text{as} \quad r \to \infty \; (l \to \pm \infty).
\end{equation}

At the throat, $r = b(r)$, the metric satisfies:

\begin{equation}
1 - \frac{b(r)}{r} \to 0.
\end{equation}

\begin{figure}
\includegraphics[width=0.7 \linewidth]{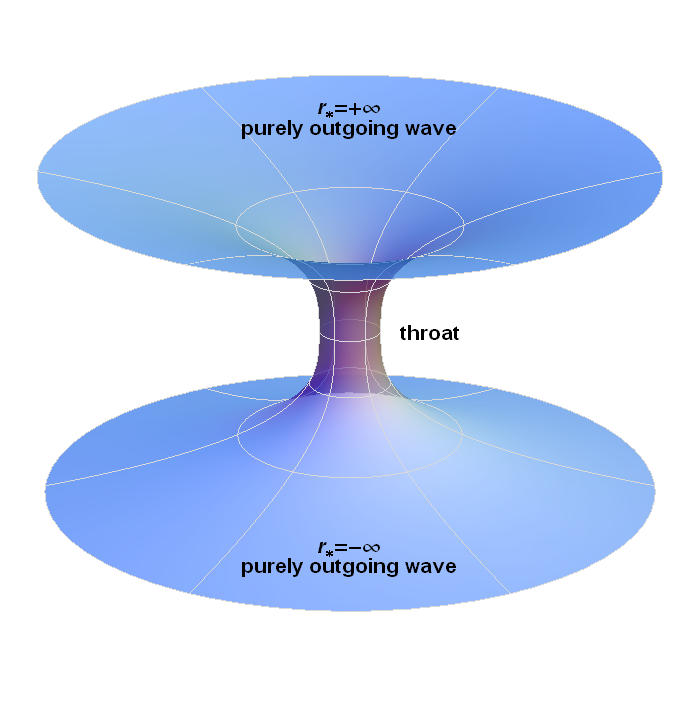}
\caption{A simple illustration of the quasinormal modes boundary conditions for a wormhole. }\label{WHfigure}
\end{figure}

The traversable wormhole metric remains non-singular at the throat, ensuring that travelers can pass through the wormhole within a finite proper time. This property makes the spacetime suitable for describing hypothetical scenarios of interstellar travel or other applications in theoretical physics.

\begin{figure*}
\includegraphics[width=0.5\linewidth]{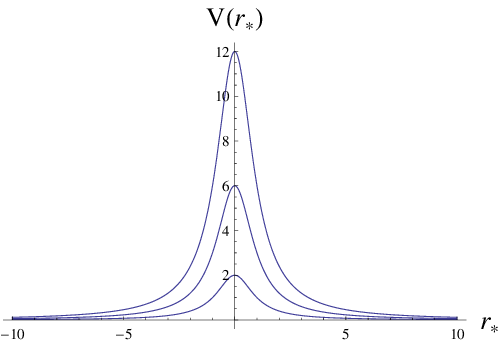}\includegraphics[width=0.5\linewidth]{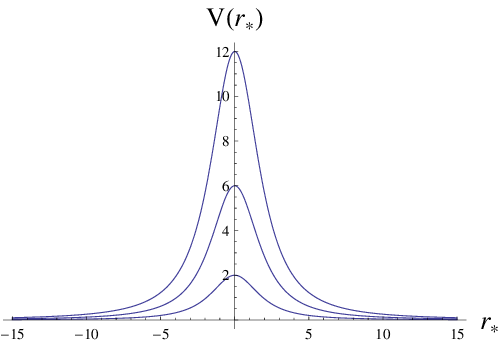}
\caption{Effective potentials for electroamgentic perturbations of the Bronnikov (left) and M-T wormholes for $\ell=1$, $2,$ and $3$ (from bottom to top), $b_{0}=1$. }\label{figPot1}
\end{figure*}

\begin{figure*}
\includegraphics[width=0.5\linewidth]{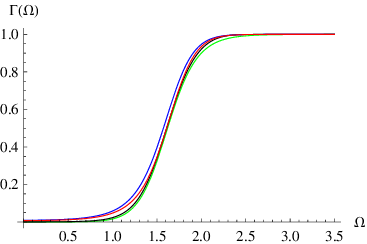}\includegraphics[width=0.5\linewidth]{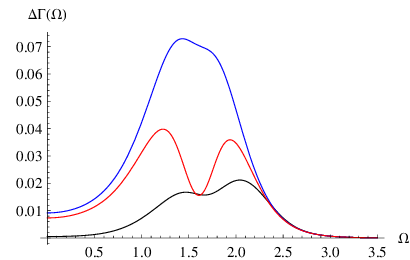}
\caption{Left panel: The grey-body factors calculated via the correspondence at various orders (eikonal - black, the first order beyond eikonal - red, and the second order beyond eikonal - blue), and the 6th order WKB method (green). Right panel: the difference between the 6th order WKB grey-body factors and those obtained via the correspondence at various orders; Here we have $\ell=1$ scalar perturbations around the Bronnikov wormhole: $b_{0}=1$.}\label{BronnikovL1s0}
\end{figure*}

\begin{figure*}
\includegraphics[width=0.5\linewidth]{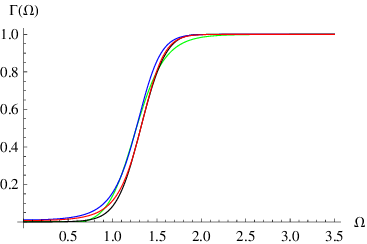}\includegraphics[width=0.5\linewidth]{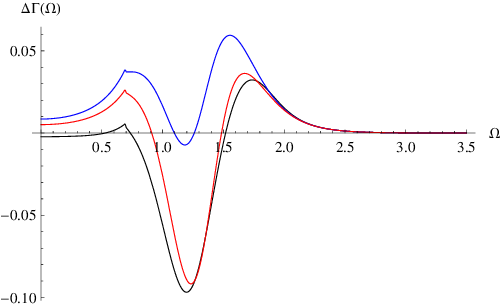}
\caption{Left panel: The grey-body factors calculated via the correspondence at various orders (eikonal - black, the first order beyond eikonal - red, and the second order beyond eikonal - blue), and the 6th order WKB method (green). Right panel: the difference between the 6th order WKB grey-body factors and those obtained via the correspondence at various orders; Here we have $\ell=1$ scalar perturbations around the Bronnikov wormhole: $b_{0}=1$.}\label{BronnikovL1s1}
\end{figure*}

\begin{figure*}
\includegraphics[width=0.5\linewidth]{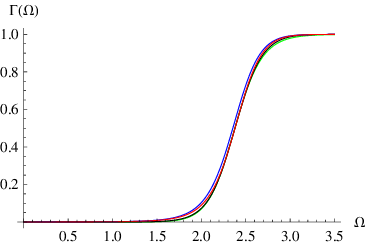}\includegraphics[width=0.5\linewidth]{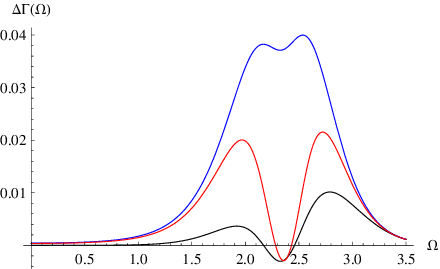}
\caption{Left panel: The grey-body factors calculated via the correspondence at various orders (eikonal - black, the first order beyond eikonal - red, and the second order beyond eikonal - blue), and the 6th order WKB method (green). Right panel: the difference between the 6th order WKB grey-body factors and those obtained via the correspondence at various orders; Here we have $\ell=2$ electromagnetic perturbations around the Bronnikov wormhole: $b_{0}=1$.}\label{BronnikovL2s1}
\end{figure*}

\begin{figure*}
\includegraphics[width=0.5\linewidth]{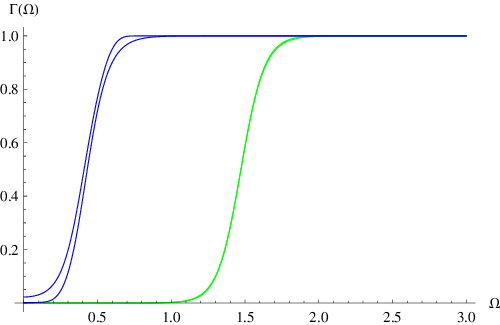}\includegraphics[width=0.5\linewidth]{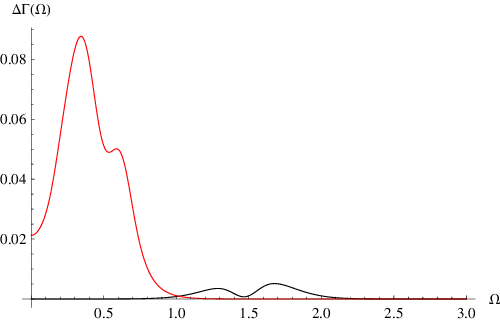}
\caption{The grey-body factors calculated via the correspondence and the 6th order WKB method (left) and the difference between them for $\ell=0$ (blue - for grey-body factors, red - for the difference) $\ell=1$ (green - for grey-body factors, black - for the difference) and  scalar  perturbations around the Morris-Thorne wormhole given by eq. \ref{MT1}: $b_{0}=1$. }\label{MTL01}
\end{figure*}

\begin{figure*}
\includegraphics[width=0.5\linewidth]{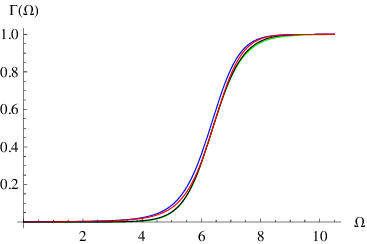}\includegraphics[width=0.5\linewidth]{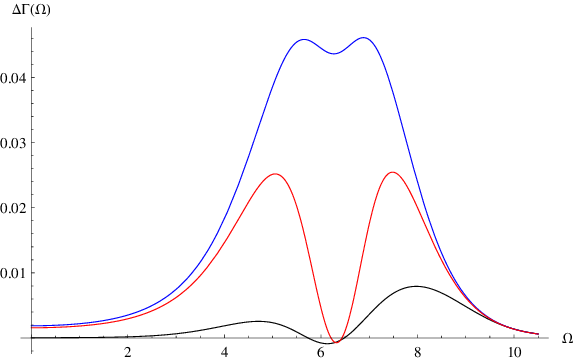}
\caption{Left panel: The grey-body factors calculated via the correspondence at various orders (eikonal - black, the first order beyond eikonal - red, and the second order beyond eikonal - blue), and the 6th order WKB method (green). Right panel: the difference between the 6th order WKB grey-body factors and those obtained via the correspondence at various orders; Here we have $\ell=2$ electromagnetic perturbations around the wormhole given by eq. \ref{MT2}: $b_{0}=1$, $p=1$, $q=0.5$.}\label{MT2L2}
\end{figure*}

\section{The correspondence}

After separating variables and performing algebraic transformations, the perturbation equations for test scalar and electromagnetic fields can be rewritten in a wave-like form for certain wave functions $\Psi$ (see, for instance, \cite{Konoplya:2006rv}):
\begin{equation}\label{sp}
\frac{d^2\Psi}{dr_*^2} + \omega^2 \Psi - V(r) \Psi = 0,
\end{equation}
where the tortoise coordinate, $r_*$, is defined as:
\begin{equation}
dr_* = \pm \frac{dr}{f(r)} = \pm \frac{dr}{e^{\Phi}\sqrt{1 - \frac{b(r)}{r}}}.
\end{equation}
The effective potential $V(r)$ is given by:
\begin{equation}
V(r) = \left(\frac{e^{2\Phi}\ell (\ell+1)}{r^2} + (s-1) \frac{f(r) f'(r)}{r}\right),
\end{equation}
where $s=0$ corresponds to the scalar field and $s=1$ corresponds to the electromagnetic field. 

For the electromagnetic field ($s=1$), the effective potential as a function of $r$ is independent of the shape function $b(r)$. However, the complete wave equation and the effective potential expressed in terms of the tortoise coordinate $r_*$ do depend on $b(r)$.  Examples of effective potentials are shown in figs. \ref{figPot1}.

In the tortoise coordinate framework, the effective potentials associated with the metric (\ref{MT}) form positive-definite potential barriers. These barriers have their maximum value located at the throat of the wormhole. 

The spacetime spans between two asymptotic regions, often referred to as "infinities," which connect either two different universes or two distant regions within the same universe. As shown in \cite{Konoplya:2005et}, the quasinormal modes (QNMs) of wormholes are solutions to the wave equation (\ref{sp}) that satisfy the following boundary conditions:
\begin{equation}
\Psi \sim e^{\pm i \omega r_*}, \quad r_* \rightarrow \pm \infty.
\end{equation}
These conditions imply purely outgoing waves at both asymptotic regions, with no incoming waves from either side (see figure \ref{WHfigure}). 

A quasinormal mode can be expressed as:
\begin{equation}
\omega = \omega_{\text{Re}} + i \omega_{\text{Im}},
\end{equation}
where $\omega_{\text{Re}}$ represents the real oscillation frequency, and $\omega_{\text{Im}}$ is proportional to the decay rate of the mode. The real part determines the oscillatory behavior, while the imaginary part governs the damping or growth of the wave amplitude over time.

The WKB method is a reliable approach for finding quasinormal modes (QNMs) in the large $\ell$ regime. This technique relies on expanding the wave function asymptotically at both infinities (event horizon and spatial infinity) and matching it with a Taylor expansion near the peak of the effective potential. The method assumes the existence of two turning points and a monotonically decreasing effective potential towards both infinities. The WKB condition is given by:
\begin{equation}\label{WKB}
\K =\frac{i Q_{0}}{\sqrt{2 Q_{0}''}} - \sum_{i=2}^{p} \Lambda_{i} = n + \frac{1}{2}, \quad n = 0, 1, 2, \ldots,
\end{equation}
where $\Lambda_i$ are correction terms, derived in works such as \cite{Schutz:1985km, Iyer:1986np, Konoplya:2003ii, Matyjasek:2017psv}, and $Q_0^{i}$ represents the $i$-th derivative of $Q = \omega^2 - V$ at the potential's maximum with respect to the tortoise coordinate $r^\star$. Here, $n$ denotes the overtone number.

The WKB approach has been extensively used and analyzed in diverse settings \cite{Bolokhov:2023ruj, Bolokhov:2023bwm, Bolokhov:2023dxq, Skvortsova:2023zmj, Skvortsova:2024wly, Konoplya:2001ji, Paul:2023eep, Zinhailo:2019rwd, Zinhailo:2018ska, Cuyubamba:2016cug, Kokkotas:2010zd, Guo:2023ivz, Gong:2023ghh,Skvortsova:2024atk,Dubinsky:2024fvi,Dubinsky:2024hmn}, including asymptotically flat spacetimes and those with asymptotic de Sitter behavior, such as the asymptotically de Sitter BTZ-like black holes studied in \cite{Konoplya:2020ibi}. While our focus is on asymptotically flat wormholes, we anticipate similar conclusions for asymptotically de Sitter wormholes.

We analyze scalar and electromagnetic perturbations, noting that gravitational perturbations, which generally begin at larger $\ell = s$ (where $s$ is the spin of the field), are usually more accurately modeled than $\ell=0$  scalar and $\ell=1$ and electromagentic perturbations using WKB data \cite{Konoplya:2019hlu,Konoplya:2024lir}.  There are certainly exceptions, because the effective potential for gravitational perturbations may have a negative gap or additional bump, producing additional turning points and making the WKB approach inaccurate (see, for instance \cite{Konoplya:2025hgp}.)  The primary advantage of the WKB method lies in its accuracy and convergence in the eikonal regime ($\ell \to \infty$). In this regime, the perturbation spectrum is often qualitatively consistent across field spins, although higher-order corrections ($1/\ell$ and beyond) can vary quantitatively between different fields.

In the study of wave scattering processes in the vicinity of black hole or wormholes, an intriguing phenomenon occurs due to the interaction of waves with the potential barrier surrounding the compact object. Waves can partially reflect off the barrier or partially transmit through it, and this process is governed by the same grey-body factors, irrespective of whether the wave originates near the black hole's event   horizon (or the asymptotic region in the other universe for wormholes) or arrives from spatial infinity. This symmetry is a fundamental feature of black hole scattering phenomena and dictates the form of the boundary conditions, which are expressed as follows:

\begin{equation}
\begin{array}{rclcl}
\Psi &=& e^{-i\Omega r_*} + R e^{i\Omega r_*}, &\quad& r_* \to +\infty, \\
\Psi &=& T e^{-i\Omega r_*}, &\quad& r_* \to -\infty,
\end{array}
\end{equation}

Here, $\Psi$ represents the wave function, $R$ is the reflection coefficient, and $T$ is the transmission coefficient. The tortoise coordinate $r_*$ maps spatial locations in such a way that the event horizon corresponds to $r_* \to -\infty$ and spatial infinity to $r_* \to +\infty$.  Notice that here we should always distinguish the real and continuous frequency $\Omega$ in the scattering problem from the complex and discrete quasinormal mode $\omega_{n}$.  

In the context of black hole physics, the transmission coefficient, $T$, plays a particularly significant role and is commonly referred to as the grey-body factor. This factor quantifies the fraction of the wave that successfully traverses the potential barrier, contributing to the emission of radiation from the black hole. Mathematically, the grey-body factor is defined as:
\begin{equation}
\Gamma_{\ell}(\Omega) = |T|^2 = 1 - |R|^2,
\end{equation}
where $\Gamma_{\ell}(\Omega)$ is the grey-body factor associated with the angular momentum number $\ell$ and frequency $\Omega$ of the wave.  The sum of the transmitted and reflected energy fractions equals unity.

The correspondence between grey-body factors and QNMs, initially established for spherically symmetric and asymptotically flat black holes, is derived using the WKB expression for grey-body factors:
\begin{equation}\label{eq:gbfactor}
\Gamma_{\ell}(\Omega) = \frac{1}{1 + e^{2\pi \imo \K}}.
\end{equation}
This WKB formula has been extensively used for finding grey-body factors \cite{Konoplya:2020jgt,Bolokhov:2024voa,Konoplya:2023ahd,Skvortsova:2024msa,Toshmatov:2015wga,Dubinsky:2024nzo}
Following the procedure outlined in \cite{Konoplya:2024lir} for black holes, we extend the analysis to wormholes, demonstrating that similar steps lead to the correspondence for wormholes as well.

Following \cite{Konoplya:2024lir},  the effective potential is expressed as:
\begin{equation}\label{potential-multipole}
V(r_*) = \ell^2 V_0(r_*) + \ell V_1(r_*) + V_2(r_*) + \ell^{-1} V_3(r_*) + \ldots.
\end{equation}
The eikonal approximation is derived from the first-order WKB formula \cite{Schutz:1985km}:
\begin{equation}\label{WKBformula-eikonal}
\Omega =   \ell \sqrt{V_{00}} - \imo \K \sqrt{\frac{-V_{02}}{2V_{00}}} + \Order{\ell^{-1}},
\end{equation}
where $V_{00}$ is the maximum value of $V_0(r_*)$, and $V_{02}$ is its second derivative at this point. This provides a reliable estimate for the dominant quasinormal modes ($n=0$, $\K = 1/2$):

\begin{equation}\label{WKBformula-eikonal-dominant}
\omega_0 = \ell \sqrt{V_{00}} - \frac{\imo}{2} \sqrt{\frac{-V_{02}}{2V_{00}}} + \Order{\ell^{-1}}.
\end{equation}
Thus, the real and imaginary parts of $\omega_0$ are:
\[
\begin{aligned}
\re{\omega_0} &= \ell \sqrt{V_{00}} + \Order{\ell^{-1}}, \\
\im{\omega_0} &= -\frac{1}{2} \sqrt{\frac{-V_{02}}{2V_{00}}} + \Order{\ell^{-1}}.
\end{aligned}
\]
The parameter $\K$ can be expressed as a function of the real frequency $\Omega$, using $\omega_0$:
$$-\imo \K = \frac{\Omega^2 - \ell^2 V_{00}}{\ell \sqrt{-2V_{02}}} + \Order{\ell^{-1}}$$
\begin{equation}\label{eikonal-K}
 = -\frac{\Omega^2 - \re{\omega_0}^2}{4 \re{\omega_0} \im{\omega_0}} + \Order{\ell^{-1}}.
\end{equation}

The transmission coefficient, linking grey-body factors $\Gamma_{\ell}(\Omega)$ with the fundamental mode $\omega_0$, is:
\begin{equation}\label{transmission-eikonal}
\Gamma_{\ell}(\Omega) \equiv |T|^2 = \left(1 + e^{2\pi \frac{\Omega^2 - \re{\omega_0}^2}{4 \re{\omega_0} \im{\omega_0}}}\right)^{-1} + \Order{\ell^{-1}}.
\end{equation}
This relationship is exact in the eikonal limit $\ell \to \infty$ and approximate for small $\ell$. We show that this connection extends beyond the eikonal regime by including overtone corrections.

Finally, taking into account all the corrections up to the third order we have: \cite{Konoplya:2024lir}
\begin{eqnarray}\nonumber
&&\imo\K=\frac{\Omega^2-\re{\omega_0}^2}{4\re{\omega_0}\im{\omega_0}}\Biggl(1+\frac{(\re{\omega_0}-\re{\omega_1})^2}{32\im{\omega_0}^2}
\\\nonumber&&\qquad\qquad-\frac{3\im{\omega_0}-\im{\omega_1}}{24\im{\omega_0}}\Biggr)
-\frac{\re{\omega_0}-\re{\omega_1}}{16\im{\omega_0}}
\\\nonumber&& -\frac{(\omega^2-\re{\omega_0}^2)^2}{16\re{\omega_0}^3\im{\omega_0}}\left(1+\frac{\re{\omega_0}(\re{\omega_0}-\re{\omega_1})}{4\im{\omega_0}^2}\right)
\\\nonumber&& +\frac{(\omega^2-\re{\omega_0}^2)^3}{32\re{\omega_0}^5\im{\omega_0}}\Biggl(1+\frac{\re{\omega_0}(\re{\omega_0}-\re{\omega_1})}{4\im{\omega_0}^2}
\\\nonumber&&\qquad +\re{\omega_0}^2\Biggl(\frac{(\re{\omega_0}-\re{\omega_1})^2}{16\im{\omega_0}^4}
\\&&\qquad\qquad -\frac{3\im{\omega_0}-\im{\omega_1}}{12\im{\omega_0}}\Biggr)\Biggr)+ \Order{\frac{1}{\ell^3}}.
\label{eq:gbsecondorder}
\end{eqnarray}
Here $\omega_0$ and $\omega_1$ are, respectively, the dominant mode and the first overtone.

\section{Testing the Correspondence: Examples}

\subsection{Bronnikov-Ellis Wormhole}

Let us first consider the case of tideless wormholes, where $\Phi = 0$. The term \textit{tideless wormhole} refers to a wormhole where the tidal forces at the throat vanish for a point particle. However, it is important to note that this condition generally does not hold for extended objects, which may still experience tidal effects. 

The Bronnikov-Ellis wormhole is a well-known example of a traversable wormhole with the following metric functions \cite{Bronnikov:1973fh,Ellis:1973yv}:
\begin{equation}
\Phi(r) = 0, \quad b(r) = \frac{b_{0}^2}{r}.
\end{equation}
\vspace{1mm}

Grey-body factors computed via the correspondence at various orders, together with the 6th-order WKB method and the differences between them, are shown in Fig.~\ref{BronnikovL1s0} for scalar perturbations with $\ell = 0$. Figs.~\ref{BronnikovL1s1} and \ref{BronnikovL2s1} display the results for electromagnetic perturbations with $\ell = 1$ and $\ell = 2$, respectively. These plots illustrate a clear convergence of the correspondence as the order of the expansion beyond the eikonal limit increases, demonstrating the robustness of the method.

\subsection{Tideless Morris-Thorne Wormhole}

Next, let us consider another example: the tideless Morris-Thorne wormhole described by Eq.~(A1) in \cite{Morris:1988cz}. In this case, the shape function is given by:
\begin{equation}\label{MT1}
b(r) = \sqrt{b_{0} r}.
\end{equation}

Grey-body factors computed via the correspondence at the second order beyond the eikonal limit and the 6th-order WKB method, along with the differences between them, are shown in Fig.~\ref{MTL01} for scalar perturbations with $\ell = 0$ and $\ell = 1$. From these results, we observe that the accuracy of the correspondence improves as $\ell$ increases, highlighting the utility of the correspondence for higher angular momentum modes.

\subsection{More General Morris-Thorne Wormholes}

Now, we consider a more general case of a Morris-Thorne wormhole where the tidal forces at the throat are non-zero. This occurs when the redshift function is no longer constant. The shape and redshift functions for this wormhole are given by:
\begin{equation}\label{MT2}
b(r) = b_0 \left(\frac{b_0}{r}\right)^q, \; q < 1, \quad \Phi(r) = \frac{1}{r^p}, \; p > 0.
\end{equation}

The scattering properties and quasinormal ringing of this wormhole have been extensively studied in \cite{Konoplya:2010kv}. As an illustrative example, Fig.~\ref{MT2L2} shows the results for electromagnetic perturbations with $\ell = 2$. These results demonstrate that including higher-order corrections to the eikonal approximation significantly enhances the accuracy of the correspondence, even for this more general class of wormholes. The analysis reaffirms the reliability of the method in capturing the essential physical features of such spacetimes.

Notice that when we refer to grey-body factors calculated at various orders beyond the eikonal correspondence, we mean that they were obtained using equation (\ref{eq:gbsecondorder}) from quasinormal modes, which were consistently computed via the sixth-order WKB approach with Padé approximants. These grey-body factors, determined at different orders of the correspondence, are then compared to the grey-body factors computed here at the sixth WKB order. The latter should be fairly close to the precise values for single-peak effective potentials, shown in Fig. \ref{figPot1}. One should not mistakenly assume that the grey-body factors were computed using the WKB formula at various orders instead.

To rigorously prove the convergence of the correspondence, we would need to determine precise values for quasinormal modes and compare them with accurate grey-body factors obtained through numerical integration. However, it is well established that quasinormal modes computed using Padé approximants at the sixth WKB order for nonzero multipole numbers typically achieve high accuracy—approximately five digits for $\ell=1$ and even greater precision for higher $\ell$ values \cite{Konoplya:2019hlu}. The same level of accuracy generally holds for grey-body factors. Meanwhile, the error due to the finite order of the correspondence may already appear at the second or third significant digit. 

Thus, although we are not strictly employing exact methods to compute quasinormal modes and grey-body factors for testing the correspondence, the relative error associated with these methods remains significantly smaller than the error introduced by the finite order of the correspondence. In other words, from a practical standpoint, demonstrating the convergence of the correspondence at a few orders beyond the eikonal limit should be sufficient. Nonetheless, it would certainly be desirable to perform precise calculations of grey-body factors and quasinormal modes for wormholes across various configurations.

\section{Conclusions}

In this work, we have demonstrated that the correspondence between quasinormal modes (QNMs) and grey-body factors (GBFs), previously established for black holes, extends naturally to a wide class of traversable wormholes. By leveraging the WKB approximation and exploring corrections beyond the eikonal limit, we verified the robustness of the correspondence using several well-known wormhole models.

Our analysis confirms that the eikonal approximation provides a reliable framework for understanding the relationship between QNMs and GBFs in wormhole spacetimes. For tideless wormholes, such as the Bronnikov-Ellis and Morris-Thorne wormholes, the correspondence converges quickly, particularly for higher angular momentum modes
$\ell$. Even for wormholes with non-zero tidal forces, where the redshift function is no longer constant, higher-order corrections to the eikonal limit improve the accuracy of the correspondence. This demonstrates the flexibility and reliability of the method for diverse wormhole geometries.

The results presented here have several implications. First, the correspondence offers a novel theoretical tool for probing wormhole spacetimes and distinguishing them from black holes based on their spectral characteristics. Second, the improved convergence beyond the eikonal limit suggests that the correspondence may be used effectively for low-
$\ell$ modes, expanding its applicability to more realistic observational scenarios.

While this study has established the validity of the QNM-GBF correspondence for spherically symmetric traversable wormholes, several open questions remain. A natural extension involves exploring the correspondence for rotating wormholes, where coupled perturbation equations and frame-dragging effects introduce additional complexity. Similarly, higher-dimensional wormholes or spacetimes with additional fields, such as scalar or dilaton fields, may provide new insights into the universality of the correspondence \cite{Bronnikov:2022bud,Bronnikov:2017sgg,Bronnikov:2016xyp}. Another direction is to investigate wormholes with asymmetric geometries relative to the throat, as the correspondence is expected to hold due to the similarity of derivation steps in terms of the tortoise coordinate. However, in some range of parameters asymmetric rotating wormholes may have a phenomenon of superradiance, similar to that observed for rotating black holes \cite{1971JETPL..14..180Z,Starobinskil:1974nkd,Starobinsky:1973aij,Bekenstein:1998nt,East:2017ovw,Hod:2012zza,Konoplya:2008hj}. The superradiance corresponds to the transmission coefficients larger than unity, and, thereby, could not be described by the current WKB approach. Consequently we expect that the correspondence should not work in this regime.

Further refinement is also needed to systematically address low-$\ell$ modes, and cases where the WKB approximation is less reliable. Finally, identifying observational signatures that distinguish wormholes from black holes, particularly in gravitational wave and electromagnetic spectra, remains an important challenge for testing these theoretical predictions.

\vspace{3mm}
\acknowledgments{
The authors would like to acknowledge R. A. Konoplya for useful discussions.
This work was supported by RUDN University research project FSSF-2023-0003.}

\bibliography{bibliography}

\end{document}